# Exact Kohn-Sham vs. Hartree-Fock in momentum-space: examples of two-fermion systems

Sébastien RAGOT

Laboratoire Structure, Propriété et Modélisation des Solides (CNRS, Unité Mixte de Recherche 85-80). École Centrale Paris, Grande Voie des Vignes, 92295 CHATENAY-MALABRY, FRANCE

## Abstract

The question of how density functional theory (DFT) compares with Hartree-Fock (HF) for the computation of momentum-space properties is addressed in relation to systems for which (near) exact Kohn-Sham (KS) and HF one-electron matrices are known. This makes it possible to objectively compare HF and exact KS and hence to assess the potential of DFT for momentum space studies. The systems considered are the Moshinsky atom, the Hooke's atom and light two-electron ions, for which expressions for correlated density-matrices or momentum densities have been derived in closed-form. The results obtained show that it is necessary to make a distinction between true and approximate DFT.

# 1. Introduction

## 1.1 Momentum-space experiments

Compton scattering and positron annihilation provide information complementary to X-ray or neutron diffraction, making it possible to investigate momentum-space electron properties of a variety of systems. In particular, X-ray Compton scattering experiments allows for probing the ground-state electronic structure of materials. A Compton scattered radiation undergoes a Doppler broadening due to the motion of electrons in the target material. The broadened signal, i.e. proportional to the directional Compton profile, is closely related to the electron momentum density $n(\mathbf{p})$ in the scatter and further depends on the scattering vector. Under impulse approximation, the directional profile $J(q,\mathbf{u})$ writes as [1,2]:

$$J(q,\mathbf{u}) = \int n(\mathbf{p})\delta(q-\mathbf{p}.\mathbf{u})d\mathbf{p} \qquad (1)$$

where $q$ stands for the momentum variable and $\mathbf{u}$ points along the scattering vector.

The nature of momentum electron distributions makes this technique particularly sensitive to the slowest electrons, whereby bonding in condensed matter or molecules can be probed. Thus, Compton experiments provide a critical test for the quantum-mechanical description of such systems, allowing for quantitative comparisons with theoretical models.

To this aim, momentum distributions may be computed from various wavefunction-based models, including sophisticated schemes, e.g. configuration interaction (CI) or the like, or more tractably Hartree-Fock (HF) calculations. In addition, there is an obvious need for computing momentum distributions from density functional theory (DFT) methods. However, as outlined by Thakkar [3], forming momentum densities from Kohn-Sham (KS) orbitals remains an unsolved though early investigated problem. This point is discussed now.

## 1.2 Definitions

Be $x_i$ the space and spin coordinates of electron $i$; the $k$-electron reduced density-matrix (or $k$RDM) derived from a $N$-electron wave function $\psi$ writes as [4]:

$$\gamma_k(x_1, x_2, ..., x_k; x_1', x_2', ..., x_k') \qquad (2)$$
$$= C_k^N \int \psi(x_1, ..., x_k, x_{k+1}, ..., x_N) \psi^*(x_1', ..., x_k', x_{k+1}, ..., x_N) dx_{k+1} dx_{k+2} .. dx_N$$

where the binomial coefficient normalizes the reduced matrix. In particular, the 1RDM is

$$\gamma_1(x_1; x_1') = N \int \psi(x_1, x_2, ..., x_N) \psi^*(x_1', x_2, ..., x_N) dx_2 ... dx_N . \qquad (3)$$

Integrating eq. (3) over spin variables leads to a "spinless" 1RDM

$$\rho_1(\mathbf{r}_1; \mathbf{r}_1') = \int [\gamma(x_1; x_1')]_{s_1 = s_1'} ds_1 . \qquad (4)$$

In the following, spinless quantities are assumed.

We may further rewrite $\rho_1(\mathbf{r}_1; \mathbf{r}_1')$ in terms of centre-of-mass and relative coordinates

$$\tilde{\rho}_1(\mathbf{R}, \mathbf{s}) \equiv \rho_1(\mathbf{r}_1; \mathbf{r}_1'), \qquad (5)$$

where $\mathbf{R}$ stands for $(\mathbf{r} + \mathbf{r'})/2$ and $\mathbf{s}$ is the difference $\mathbf{r} - \mathbf{r'}$, which measures some "delocalization" of particles.

The electron charge density reduces to $\rho(\mathbf{R}) = \tilde{\rho}_1(\mathbf{R}, 0) = \rho_1(\mathbf{R}; \mathbf{R})$; information on $\mathbf{s} = \mathbf{r} - \mathbf{r'}$ is thus lost.

Next, the momentum density is defined [1,4] as

$$n(\mathbf{p}) = \frac{1}{(2\pi)^3} \int \rho_1(\mathbf{r}; \mathbf{r'}) e^{i\mathbf{p} \cdot (\mathbf{r} - \mathbf{r'})} d\mathbf{r} d\mathbf{r'} = \frac{1}{(2\pi)^3} \int \tilde{\rho}_1(\mathbf{R}, \mathbf{s}) e^{i\mathbf{p} \cdot \mathbf{s}} d\mathbf{R} d\mathbf{s} . \qquad (6)$$

Thus, the computation of $n(\mathbf{p})$ requires projecting the $N$-electron wavefunction in the one-particle subspace, while keeping information related to $\mathbf{s} = \mathbf{r} - \mathbf{r'}$. There is therefore no direct route from $\rho(\mathbf{R})$ to $n(\mathbf{p})$, though $n(\mathbf{p})$ could in principle be formulated as a functional of the density $\rho(\mathbf{R})$, as any electronic property of the $N$-electron system [5].

Rather, the computation of $n(\mathbf{p})$ from direct space requires in practice at least information on $\mathbf{s} = \mathbf{r} - \mathbf{r'}$; $n(\mathbf{p})$ can thus be obtained from the auto correlation function $B(\mathbf{s})$, implicitly defined in eq. (6) as:

$$B(\mathbf{s}) = \int \tilde{\rho}_1(\mathbf{R},\mathbf{s})d\mathbf{R} . \tag{7}$$

Incidentally, it follows from eq. (6) that $B(\mathbf{s})$ fulfils the normalization condition

$$B(\mathbf{0}) = \int \tilde{\rho}_1(\mathbf{R},\mathbf{0})d\mathbf{R} = \int \rho(\mathbf{R})d\mathbf{R} = \int n(\mathbf{p})d\mathbf{p} = N . \tag{8}$$

Also, various momentum-space (or **P**-space) properties can be computed directly from $B(\mathbf{s})$, including the kinetic energy $T = -\frac{1}{2}\nabla_s^2 B(s)\big|_{s=0} = \int \frac{1}{2} p^2 n(\mathbf{p})d\mathbf{p}$ and, as an alternative to eq. (1), the directional impulse Compton profiles, i.e. $J(q,\mathbf{u}) = \frac{1}{2\pi}\int B_1(s\mathbf{u})e^{iqs}ds$.

In practice, standard *ab initio* programs provide matrix elements $\rho_{ij}^{AB}$ of any one-electron matrix (CI, HF, KS, etc.), in a basis of atomic orbitals $\varphi_{iA}$ (the $i^{th}$ orbital of atom $A$). From the expression

$$\rho_1(\mathbf{r},\mathbf{r'}) = \sum_{A,B} \sum_{i \in A, j \in B} \rho_{ij}^{AB} \varphi_{iA}(\mathbf{r} - \mathbf{R}_A) \varphi_{jB}(\mathbf{r'} - \mathbf{R}_B), \tag{9}$$

where $R_A$ points at nucleus of atom $A$, it is therefore possible to reconstruct $n(\mathbf{p})$ according to eq. (6).

### 1.3 Dilemma

Basically, there is a need for computing momentum densities directly from a DFT calculation, with sufficient confidence in the accuracy for systematic quantitative comparison with experiments. However, computing the momentum density directly from a KS-1RDM (hereafter $n_{KS}(\mathbf{p})$), is questionable in the extent that $n_{KS}(\mathbf{p})$ reflects the associated KS determinant corresponding to the fictious non-interacting system involved in this theory [5]. The resulting $n_{KS}(\mathbf{p})$ does hence not reflect the true wavefunction of the system. Thus, even if exact exchange-correlation energy were available (hence giving exact energies), the subsequent $n_{KS}(\mathbf{p})$ would not be consistent with the exact DFT kinetic energy as part of it resides outside the KS determinant [5]. In contrast, a momentum density computed from a 1RDM derived from a model wavefunction unambiguously reflects said wavefunction. This is in particular true at HF level, so that HF if sometimes seen as a more suitable theory for **P**-space studies.

Incidentally, we keep in mind that it is not the purpose of DFT to be accurate in **P**-space (this is not required in this theory).

Yet, the scheme proposed by Lam and Platzman [6] is known to include correlation corrections to momentum densities within DFT. In the context of the local density approximation (LDA), these authors have proposed a practical scheme, in which the correction to the momentum density is computed from the difference between the interacting and non-interacting momentum densities, respectively $n_I(\mathbf{p})$ and $n_{NI}(\mathbf{p})$, taken as local functionals of the density $\rho(\mathbf{R})$ of the system considered, that is

$$\Delta n(\mathbf{p}) = \int \{n_I[\rho](\mathbf{p}) - n_{NI}[\rho](\mathbf{p})\}\rho(\mathbf{r})d\mathbf{r} . \tag{10}$$

This approach, isotropic by construction, has been used extensively, mostly for metals and semiconductors, which is natural in the context of the LDA. One may for instance refer to ref. [7] for discussion and comparison of the above scheme with an alternative including anisotropic corrections. The Lam-Platzman scheme has further been applied to atomic systems, bringing KS results close to HF ones [6,8], though numerical difficulties may occur due to long atomic density tails.

In practice, corrections to KS momentum densities or Compton profiles are however often bypassed. The meaning of the resulting quantities is therefore questionable, as outlined in ref. [9]. In addition, the quality of the results strongly depends on the functionals used. It follows that it is unclear whether a KS approach is better than HF for **P**-space or not. For example, it was concluded that KS and HF approaches were comparable, either assessing approximate Dyson orbitals for the calculation of electron-momentum-spectroscopy scattering cross sections [10] or Compton profiles of water and mixed water-neon clusters [11]. Other studies have claimed superiority of HF for **P**-space studies of ice [12].

The systematic investigation by Thakkar [3] of a set of 319 singlet state molecules helped in drawing more general conclusions. In this work, the calculation of **P**-space properties from KS matrices turned on average worse than HF, at least when using popular DFT approaches such as B3LYP within cc-PVTZ basis-set.

It would be unfair to extend such conclusions to any DFT approach, due to the diversity of DFT methods (the DFT "zoology"). Indeed, aside the most popular DFT approaches, one may contemplate using functionals suitable for **P**-space studies. In this spirit, an investigation of Zope [13] has shown that a *self-interaction corrected* LDA formalism substantially improves atomic Compton profiles with respect to generalized gradient corrected (GGA) approaches. We note here that implementing self-interaction corrections brings the resulting KS-1RDM closer to HF-1RDM as the latter theory also excludes self-interactions [5], which may explain the obtained improvement [14].

Further, in a different spirit, Harbola and co-workers [15] proposed a scheme valid for arbitrary external potentials to compute **P**-space properties, based on a variant [16] of the Levy's constrained search approach [17] and using accurate **R**-space electron densities. Application to atomic systems He, Be and Ne has led to results slightly better than HF on average when compared with more accurate theories.

The DFT momentum moments of He, Be and Ne obtained from the methods used by Zope, Harbola and co-workers [13,15] are tabulated in table 1, compared with their HF [18] and correlated wavefunction counterparts [19,20,21]. As can be seen in this table, the momentum moments of refs. [13,15] may improve over HF values, especially at high momenta (considering correlated wavefunction methods as a reference). Thus, it seems that there is a room for at least reaching HF-like accuracy when computing Compton profiles from KS orbitals, even if corrections are bypassed.

In this respect, one may point out that though there is no need for the KS-1RDM to be closer than HF to the exact one, it is a priori not excluded that a KS determinant leads to a more accurate momentum density. Indeed, the HF wavefunction matches a hamiltonian which decomposes as the sum of one-electron operators, that is, not the true hamiltonian. Hence, HF and KS are just two different paradigms involving determinants.

Now, though the HF theory minimizes the determinantal energy, nothing prevents a different determinant (which by definition gives a non-minimal energy) to lead to a higher kinetic energy, as expected from the virial theorem for Coulomb systems. Therefore, said different determinant may possibly leads to a more accurate momentum density, especially at high momenta (this is what happens in table 1, when using the approaches of refs. [13,15]).

Moreover, since the KS scheme allows in principle to reach the exact **R**-space density, in contrast with HF, one may decently hope recovering more accurate **P**-space properties from the KS determinant only, even if no correction is contemplated.

Therefore, the natural first step to assess the potential of DFT versus HF for momentum-space studies would be to compare *exact* HF, *exact* KS and (near) *exact* results, at least for those systems making it is possible.

To this aim, we investigate here a variety of two-fermion systems, for which exact KS orbitals can be formulated [22]. More specifically, **P**-space properties of the so-called Moshinsky atom [4,23,24], the Hooke's atom [22] and light two-electron systems (H$^-$, He) are compared under various approximations in the next section (sect. II). Considerations on the 1RDM and auto correlation functions follow (section III) in view of a tentative conclusion.

## 2. Near exact, HF and KS momentum densities of some two-fermion systems

We can take advantage of dealing with two-particle systems to define the unique ground-state, nodeless KS orbital involved as $\varphi_{KS}(r) = \sqrt{\rho(r)/2}$, as suggested in refs. [9,16,22] (adding a phase to $\varphi_{KS}(\mathbf{r})$ would lead to higher kinetic energies [16]). Thus, starting from an accurate (possibly exact) density $\rho(\mathbf{r})$, we can calculate the Fourier transform $\chi_{KS}$ of $\varphi_{KS}$ and then the KS momentum density $n_{KS}(\mathbf{p}) = 2|\chi_{KS}(\mathbf{p})|^2$, as in ref. [9].

The corresponding HF expression is similar, e.g. $n_{HF}(\mathbf{p}) = 2|\chi_{HF}(\mathbf{p})|^2$, except that the square of HF orbital is not proportional to the exact charge density but rather slightly deviates from it (typically in relative error of a few percents).

The above expressions reflect the fact that a determinantal 1RDM can be written as a finite sum of orbital products, e.g. only one in the present case. In contrast, the exact 1RDM is known to involve an infinite sum of terms, reflecting that **r** and **r'** variables are not separable [4].

### 2.1 Moshinsky atom

This point can be easily understood through a simple example: the Moshinsky atom, in which two fermions are bounded from a harmonic potential $\frac{1}{2}k(r_1^2 + r_2^2)$ while repelling each other via the Hooke's law $-\frac{1}{2}lr_{12}^2 = -\frac{1}{2}l(\mathbf{r}_1 - \mathbf{r}_2)^2$. Such a system is exactly solvable, as for example illustrated in refs. [4,23], allowing for comparison between exact, HF and KS-like density matrices, respectively denoted by $\rho_1$, $\rho_{1,HF}$ and $\rho_{1,KS}$. Expressions obtained are respectively

$$\rho_1 \propto e^{-(\alpha R^2 + \frac{1}{4}\beta s^2)} = e^{-\frac{1}{4}(\alpha+\beta)(r^2+r'^2) - \frac{1}{2}(\alpha-\beta)\mathbf{r}\cdot\mathbf{r}'},$$

$$\rho_{1,KS} \propto e^{-\alpha\left(R^2+\frac{1}{4}s^2\right)} = e^{-\alpha\left(r^2+r'^2\right)/2}, \text{ and}$$

$$\rho_{1,HF} \propto e^{-\gamma\left(R^2+\frac{1}{4}s^2\right)} = e^{-\gamma\left(r^2+r'^2\right)/2},$$

where $\alpha$, $\beta$, and $\gamma$ are constants defined as:

$$\alpha = \frac{2\sqrt{k(k-2l)}}{\left(\sqrt{k}+\sqrt{(k-2l)}\right)},$$

$$\beta = \frac{\sqrt{k(k-2l)}+k-l}{\left(\sqrt{k}+\sqrt{(k-2l)}\right)} \text{ and}$$

$$\gamma = \sqrt{(k-l)}.$$

Some comments are in order. The exact energy $E = \frac{3}{2}\left(\sqrt{k}+\sqrt{(k-2l)}\right)$ requires $k > 2l$ for the particles to remain bounded, while the HF solution $E_{HF} = 3\sqrt{(k-l)}$ imposes $k > l$. Further, any density derived from the 1RDMs above will differ from each other by a simple scale factor in this case. Still, since $\alpha \neq \beta$ (unless $l = 0$), the exponent in $\rho_1$ is not proportional to $r^2 + r'^2$; the exact $\rho_1$ can thus not be rewritten as a single orbital product, in contrast with $\rho_{1,HF}$ and $\rho_{1,KS}$. Rather, it may be expanded as an infinite sum of orbital products, as said above; this can for example be seen by expanding the term $e^{-\frac{1}{2}(\alpha-\beta)\mathbf{r}\cdot\mathbf{r}'}$ and rearranging the expanded terms as products of single variable orbitals.

The resulting exact, HF and KS radial momentum densities are compared in fig. 1, considering the case $k = 1/4$, $l = 1/12$, where it appears that $n_{HF}(p)$ is slightly closer to $n(p)$ than $n_{KS}(p)$ in this case.

The HF and KS kinetic energies are 0.612372 and 0.549038 a.u., that is, in error of +3.5% and -7.2%, respectively, with respect to the exact kinetic energy (0.591506 a.u.). Thus, while the virial theorem implies the kinetic energy of a harmonic system in its ground state to be minimal, the KS kinetic energy obtained is less than the exact one, illustrating that part of the exact kinetic energy resides outside the KS determinant.

## 2.2 Hooke's atom

In the Hooke's atom, the electrons are still bounded from the harmonic potential $\frac{1}{2}k(r_1^2 + r_2^2)$ but now mutually interact via the Coulomb potential $1/r_{12}$. Using $k = 1/4$ allows the wavefunction and the **R**-space (charge) density to be formulated in closed-form [22,25]. Besides, the wavefunction can be formulated in momentum space [26]. Taking 2-particles momentum variables $\mathbf{P}_{12} = \mathbf{p}_1 + \mathbf{p}_2$ and $\mathbf{p}_{12} = (\mathbf{p}_2 - \mathbf{p}_1)/2$ associated to $\mathbf{R}_{12} = (\mathbf{r}_1 + \mathbf{r}_2)/2$ and $\mathbf{r}_{12} = \mathbf{r}_2 - \mathbf{r}_1$, respectively, the momentum wavefunction can be separated as $\psi = \psi_{P_{12}}(P_{12})\psi_{p_{12}}(p_{12})$, with:

$$\psi_{P_{12}}(P_{12}) = \frac{1}{\pi^{3/4}} e^{-P_{12}^2/2} \text{ and}$$

$$\psi_{p_{12}}(p_{12}) = \frac{4e^{-2p_{12}^2}}{\pi(16 + 10\pi^{1/2})^{1/2}} \left\{ (2\pi)^{1/2} + 2e^{2p_{12}^2}\left(1 - \frac{e^{-2p_{12}^2}\sqrt{\pi/2}(4p_{12}^2 - 1)\text{erfi}(\sqrt{2}p_{12})}{2p_{12}}\right) \right\},$$

where erf*i* is the imaginary error function erf(*ix*)/*i*.

Next, the momentum density can be calculated thanks to the relation:

$$n(\mathbf{p}_1) = \int \left| \psi_{P_{12}}(P_{12})\psi_{p_{12}}(p_{12}) \right|^2 d\mathbf{p}_2. \tag{11}$$

However, as a direct integration is not feasible, due to the second term of $\psi_{p_{12}}$, one may use the following expansion:

$$\left(1 - \frac{e^{-2p_{12}^2}\sqrt{\pi/2}(4p_{12}^2 - 1)\text{Erfi}(\sqrt{2}p_{12})}{2p_{12}}\right) = 2e^{-2p_{12}^2} \sum_{m=0}^{\infty} C_m p_{12}^{2m}. \tag{12}$$

where the gaussian in the right-hand term ensures the wavefunction to remain finite integrable and the coefficients $C_m$ are such that expansion coefficients of left and right-hand terms are identical to any order. Then, in eq. (11), replacing $\frac{1}{2}\mathbf{P}_{12} = \mathbf{p}_1 + \mathbf{p}_{12}$ and integrating over $d\mathbf{p}_2 \equiv d\mathbf{p}_{12}$ leads to the following expression for the exact momentum density:

$$n(p) = \frac{K_1^2 e^{-2p^2}}{2^{1/2}}$$
$$+ \frac{1}{\pi^{1/2}} K_1^2 K_2 \sum_{m=0}^{\infty} C_m 2^{\frac{3}{2}-3m} \Gamma\left(m+\frac{3}{2}\right) {}_1F_1\left(m+\frac{3}{2},\frac{3}{2},2p^2\right) e^{-4p^2}$$
$$+ \frac{1}{\pi^{1/2}} K_1^2 K_2^2 \sum_{m=0}^{\infty}\sum_{n=0}^{\infty} C_m C_n 2^{\frac{1}{2}-3(m+n)} \Gamma\left(m+n+\frac{3}{2}\right) {}_1F_1\left(m+n+\frac{3}{2},\frac{3}{2},2p^2\right) e^{-4p^2}$$

(13)

where $K_1 = \dfrac{4}{\sqrt{8\pi + 5\pi^{3/2}}}$, $K_2 = 2\sqrt{\dfrac{2}{\pi}}$, $\Gamma$ is the Euler gamma function and ${}_1F_1$ is the Kummer confluent hypergeometric function. Such an expansion of $n(p)$ converges rather slowly, e.g. as a result of the non-separability of the $\mathbf{r}$ and $\mathbf{r}'$ variables in the 1RDM (see table 2).

The HF momentum density was obtained by expanding the HF orbital over the harmonic-oscillator eigenfunctions, following ref. [26]. A similar expansion has been used for approximating the KS orbital, except that the coefficients of the expansion were afterwards fitted to the exact orbital $\varphi_{KS}(\mathbf{r}) = \sqrt{\rho(\mathbf{r})/2}$, allowing momentum densities to be obtained in closed form. Using the converged value $T = 0.664418$ a.u, the difference of kinetic energies with respect to HF and KS values were found to be 0.031884 and 0.028864 (numerical integration), respectively. Thus, a small part of the quantum-chemistry-like correlation kinetic energy is picked-up by KS orbitals in that case. Accordingly, $n_{KS}(p)$ is found to be slightly closer than $n_{HF}(p)$ to the expanded $n(p)$ of eq. (13), as can be seen from the radial differences plotted in fig. 2. In contrast with the Moshinsky's atom, both $n_{KS}(p)$ and $n_{HF}(p)$ appear here to overestimate the number of momentum electrons at low momenta, consistently with the fact that kinetic energy is slightly underestimated in both cases. Additional comments are made in the next sections.

## 2.3 Case of light two-electron ions

Amongst two-electron ions, the case of the anion H⁻ is of special interest, due to the competition between inter electronic repulsion and attractive electron-nucleus potential. The HF theory does for instance predict it unstable, in contrast with accurate methods. Therefore, correlation plays a critical role.

One may for instance contemplate using the following trial ground-state wavefunction

$$\psi(r_1, r_2) = \sum_{\mu,\nu} c_{\mu\nu} \phi_{1s,\mu}(r_1) \phi_{1s,\nu}(r_2) f_{\mu\nu}(r_1, r_2), \tag{14}$$

where $\phi_{1s,\mu}(r_1)\phi_{1s,\nu}(r_2) = e^{-\mu r_1 - \nu r_2}$ accounts for radial or in-out correlation (if $\mu \neq \nu$) while the term $f_{\mu\nu}(r_1.r_2)$, here taken as $(1 - b_{\mu\nu}\, r_1.r_2)$, angularly correlates the electron pair [27]. Such a wavefunction has a simple momentum-space counterpart

$$\Psi(\mathbf{p}_1, \mathbf{p}_2) = \sum_{\mu,\nu} c_{\mu\nu} \left(1 + \frac{16 b_{\mu\nu}\, \mathbf{p}_1 \cdot \mathbf{p}_2}{(p_1^2 + \mu^2)(p_2^2 + \nu^2)}\right) \left[\frac{8\mu\nu}{\pi(p_1^2 + \mu^2)^2 (p_2^2 + \nu^2)^2}\right],$$

and further allows closed-form expressions to be derived for the 1RDM and momentum density, namely

$$\rho_1(\mathbf{r}_1, \mathbf{r}_1') = 2 \sum_{\alpha,\beta,\delta,\gamma} c_{\alpha\beta} c_{\delta\gamma} \left\{\frac{8(\alpha\beta\delta\gamma)^{3/2}\left[4 b_{\alpha\beta} b_{\delta\gamma} \mathbf{r}_1.\mathbf{r}_1' + (\beta+\gamma)^2\right] e^{-(\alpha r_1 + \delta r_1')}}{\pi(\beta+\gamma)^5}\right\}, \tag{15}$$

and

$$n(p) = 2 \sum_{\alpha,\beta,\delta,\gamma} c_{\alpha\beta} c_{\delta\gamma} \left\{\frac{64\alpha\beta(\alpha\beta\delta\gamma)^{3/2}\left[64 b_{\alpha\beta} b_{\delta\gamma} p^2 + (\beta+\gamma)^2(p^2+\alpha^2)(p^2+\delta^2)\right]}{\pi^2(\beta+\gamma)^5 (p^2+\alpha^2)^3 (p^2+\delta^2)^3}\right\}. \tag{16}$$

As can be seen from eq. (15), the obtained 1RDM shows explicit dependence between **r** and **r'** variables, via the scalar product, which indirectly results from two-particle angular correlation, eqs. (3), (4) and (14).

The correlated distributions have been computed for H⁻ from the variationally optimized wavefunction of eq. (14), allowing 93% of the exact correlation energy to be recovered and leading to a correlation Compton profile in very good agreement with the benchmark result of ref. [19], see ref. [27] for details.

Again, the KS momentum distribution has been computed from the orbital $\varphi_{KS}(r) = \sqrt{\rho(r)/2}$, where $\rho(\mathbf{r}) = \rho_1(\mathbf{r},\mathbf{r})$ in eq. (15).

The correlated, HF and KS radial momentum densities of H⁻ are compared in fig. 3. It is striking that $n_{KS}(p)$ is closer to $n(p)$ than $n_{HF}(p)$ in that case. The numerically computed difference of kinetic energies is 0.00791 a.u. (KS – HF, numerical integration) vs. 0.03815 a.u. (Correlated – HF), revealing a lack of momentum transfer in both cases. Yet, a small part of the correlation kinetic energy is recovered from the KS orbital, as in the Hooke's atom.

The same observation holds for He (not shown here, see ref. [9]). The wavefunction of eq. (14) turns however substantially less accurate in this case, compared with H⁻, due to an insufficient angular expansion of $f_{\mu\nu}(r_1,r_2)$. Yet, the KS momentum density obtained yields momentum moments in close agreement with the results of ref. [15]. The momentum moments of H- and He atomic systems are reported in table 3 together with those of Hooke's atom obtained as described above.

In line with the literature, the above results show that an exact KS matrix may lead to quite accurate **P**-space properties for 2-particles systems. In the case of the pure harmonic system (Moshinsky atom), HF gives a better momentum density. When electrons interact via the coulomb potential (Hooke's atom, H⁻, He), the exact KS momentum density is just slightly more accurate than HF.

However, the problem turns very different when the exact density is unknown, that is, when approximate functionals are needed to calculate the KS orbitals. As an example (not reported in figures), we have calculated CISD, HF and KS densities for H⁻ and He within Aug-cc-pV5Z basis-set. For the DFT calculation, use was made of two classical packages of functionals, namely (i) the Slater exchange with coefficient 2/3 [28] and the Perdew local correlation functional [29] (abbreviated as SPL) and (ii) the BPW91, that is, the Becke's 88 exchange functional [30] and Perdew/Wang's 1991 correlation functional [31]. While all computed KS charge and momentum densities substantially improve the HF ones in the case of H$^-$, the KS charge and momentum densities obtained for He reveals instead clearly less accurate (with either SPL or BPW91 scheme).

Thus, a distinction should be made between exact DFT and approximate DFT. As illustrated in the above examples, using *exact* KS orbitals slightly improves the HF computational scheme for coulombic systems. In contrast, *approximate* DFT may lead to results worse than HF's, hence recommending being cautious when directly interpreting KS orbitals in momentum space, in agreement with the conclusions drawn in refs. [3,13] (see also table 1).

## 3. Comparison of correlated, KS and HF spinless 1RDMs

### 3.1 Case of 2-electron system

Consider now a general 2-electron system with an antisymmetric spin function, the exact ground-state space wavefunction of which can formally be written as

$$\psi(\mathbf{r}_1, \mathbf{r}_2) = \psi_{HF}(\mathbf{r}_1, \mathbf{r}_2) g(\mathbf{r}_1, \mathbf{r}_2) \\ = \varphi(\mathbf{r}_1)\varphi(\mathbf{r}_2) g(\mathbf{r}_1, \mathbf{r}_2),$$  (17)

where $\varphi$ is the single HF orbital involved, assumed real and positive, and *g* correlates electrons so as to recover the exact wavefunction (*g* includes a normalization factor). In particular, the correlation factor $g(\mathbf{r}_1,\mathbf{r}_2)$ redistributes electron 1 about electron 2 along the $\mathbf{r}_{12}$ axis and one can expect it to be less than 1 for small $r_{12}$ separation, so as to decrease the probability of finding electrons close to each other. From this wavefunction, an expression for the exact 1RDM is

$$\rho(\mathbf{r}_1, \mathbf{r}_1') = 2\varphi(\mathbf{r}_1)\varphi(\mathbf{r}_1')\langle g(\mathbf{r}_1, \mathbf{r}_2) g(\mathbf{r}_1', \mathbf{r}_2)\rangle_{\mathbf{r}_2},$$  (18)

where the average denotes integration over $\mathbf{r}_2$, weighted by $\varphi^2(\mathbf{r}_2)$. Notice that the above average "correlates" $\mathbf{r}_1$ and $\mathbf{r}_1'$ indirectly through $\mathbf{r}_2$. The HF-1RDM is simply

$$\rho_{HF}(\mathbf{r}_1,\mathbf{r}_1') = 2\varphi(\mathbf{r}_1)\varphi(\mathbf{r}_1'). \tag{19}$$

Provided that the HF matrix is a good approximation to the exact matrix, the corrective term $\langle g(\mathbf{r}_1,\mathbf{r}_2)g(\mathbf{r}_1',\mathbf{r}_2)\rangle_{\mathbf{r}_2}$ should remain close to 1. However, since the factor $g(\mathbf{r}_1,\mathbf{r}_2)$ is likely to oscillate along the $\mathbf{r}_{12}$ axis, the difference $\Delta\rho(\mathbf{r}_1,\mathbf{r}_1') = \rho(\mathbf{r}_1,\mathbf{r}_1') - \rho_{HF}(\mathbf{r}_1,\mathbf{r}_1')$ has no constant sign, as exemplified in fig. 4 (left) for H$^-$. Accordingly, the corresponding difference $\Delta B(s) = B(s) - B_{HF}(s)$ is likely to show one or more node(s), as in fig. 5.

As before, the exact density $\rho(\mathbf{r}_1) = 2\varphi(\mathbf{r}_1)^2 \langle g(\mathbf{r}_1,\mathbf{r}_2)^2 \rangle_{\mathbf{r}_2}$ determines the exact KS orbital $\varphi_{KS}(\mathbf{r}_1) = \varphi(\mathbf{r}_1)\langle g(\mathbf{r}_1,\mathbf{r}_2)^2 \rangle_{\mathbf{r}_2}^{1/2}$. Thus, the corresponding KS matrix, expressed in terms of the HF orbital, is simply

$$\begin{aligned}\rho_{KS}(\mathbf{r}_1,\mathbf{r}_1') &= 2\varphi_{KS}(\mathbf{r}_1)\varphi_{KS}(\mathbf{r}_1') \\ &= 2\varphi(\mathbf{r}_1)\varphi(\mathbf{r}_1')\langle g(\mathbf{r}_1,\mathbf{r}_2)^2\rangle_{\mathbf{r}_2}^{1/2}\langle g(\mathbf{r}_1',\mathbf{r}_2)^2\rangle_{\mathbf{r}_2}^{1/2} \\ &= \rho_{HF}(\mathbf{r}_1,\mathbf{r}_1')\langle g(\mathbf{r}_1,\mathbf{r}_2)^2\rangle_{\mathbf{r}_2}^{1/2}\langle g(\mathbf{r}_1',\mathbf{r}_2)^2\rangle_{\mathbf{r}_2}^{1/2}\end{aligned} \tag{20}$$

This last equation shows how post-HF correlation is made "independent electron-like" in the KS one-matrix, whereas in the exact case, variables $\mathbf{r}_1$ and $\mathbf{r}_1'$ are not separable.

Unlike in the HF case, the difference

$$\begin{aligned}\Delta\rho(\mathbf{r}_1,\mathbf{r}_1') &= \rho(\mathbf{r}_1,\mathbf{r}_1') - \rho_{KS}(\mathbf{r}_1,\mathbf{r}_1') \\ &= 2\varphi(\mathbf{r}_1)\varphi(\mathbf{r}_1')\left\{\langle g(\mathbf{r}_1,\mathbf{r}_2)g(\mathbf{r}_1',\mathbf{r}_2)\rangle_{\mathbf{r}_2} - \langle g(\mathbf{r}_1,\mathbf{r}_2)^2\rangle_{\mathbf{r}_2}^{1/2}\langle g(\mathbf{r}_1',\mathbf{r}_2)^2\rangle_{\mathbf{r}_2}^{1/2}\right\}\end{aligned} \tag{21}$$

is clearly negative or zero owing to the Schwartz inequality ($\varphi$ is assumed real and positive). One can thus expect $\Delta B(s) = B(s) - B_{KS}(s)$ to be negative or zero for all $s$ values. Since both $B$ and $B_{KS}$ must decrease quadratically near the origin, it follows that the KS (determinantal) kinetic energy is less than the exact one, owing to $T = -\frac{1}{2}\nabla_s^2 B(s)\big|_{s=0}$.

Now, as regards the difference between exact and HF kinetic energy, a distinction should be made between Coulomb and harmonic systems, for which the virial theorem acts differently.

In the case of Coulomb systems, the exact kinetic energy must be greater than the HF one, according to the virial theorem. Thus, one expects the difference $n(p) - n_{HF}(p)$ to be positive at large $p$. The same holds for $n(p) - n_{KS}(p)$, due to eq. (21). Here the respective behaviors of $n(p) - n_{HF}(p)$ and $n(p) - n_{KS}(p)$ are much the same.

In contrast, for harmonic systems, the exact kinetic energy must be less than the HF one, owing to the virial theorem. Consistently, the difference $n(p) - n_{HF}(p)$ is likely to be negative at large $p$. However, $n(p) - n_{KS}(p)$ is expected to be positive at large $p$, due to eq. (21). Thus, the respective behaviors of $n(p) - n_{HF}(p)$ and $n(p) - n_{KS}(p)$ are opposite in this case.

These remarks consistently reflect in the various trends observed from figures 1 to 3.

Next, at large $\mathbf{r}_1$ and $\mathbf{r}_1'$ values (that is, larger than a characteristic size of $\varphi^2(\mathbf{r}_2)$), $g(\mathbf{r}_1,\mathbf{r}_2)$ should become rather insensitive to fluctuations of $\mathbf{r}_2$, so that one may approximate it by $g(\mathbf{r}_1,0)$. Thus, one has

$$\lim_{\mathbf{r}_1,\mathbf{r}_1' \to \infty}\left\{\langle g(\mathbf{r}_1,\mathbf{r}_2)g(\mathbf{r}_1',\mathbf{r}_2)\rangle_{\mathbf{r}_2} - \langle g(\mathbf{r}_1,\mathbf{r}_2)^2\rangle_{\mathbf{r}_2}^{1/2}\langle g(\mathbf{r}_1',\mathbf{r}_2)^2\rangle_{\mathbf{r}_2}^{1/2}\right\}$$
$$\approx \langle g(\mathbf{r}_1,0)g(\mathbf{r}_1',0)\rangle_{\mathbf{r}_2} - \langle g(\mathbf{r}_1,0)^2\rangle_{\mathbf{r}_2}^{1/2}\langle g(\mathbf{r}_1',0)^2\rangle_{\mathbf{r}_2}^{1/2}$$

and

$$\langle g(\mathbf{r}_1,0)g(\mathbf{r}_1',0)\rangle_{\mathbf{r}_2} - \langle g(\mathbf{r}_1,0)^2\rangle_{\mathbf{r}_2}^{1/2}\langle g(\mathbf{r}_1',0)^2\rangle_{\mathbf{r}_2}^{1/2}$$
$$= g(\mathbf{r}_1,0)g(\mathbf{r}_1',0) - |g(\mathbf{r}_1,0)||g(\mathbf{r}_1',0)|$$

So, provided that the sign of $g$ remains constant above sufficiently large values of $r_1/r_{1'}$, the difference $\rho(\mathbf{r}_1,\mathbf{r}_1') - \rho_{KS}(\mathbf{r}_1,\mathbf{r}_1')$ is likely to tend to zero faster than $\rho(\mathbf{r}_1,\mathbf{r}_1') - \rho_{HF}(\mathbf{r}_1,\mathbf{r}_1')$.

Hence, the KS matrix improves over HF at large distances. Meanwhile, the KS matrix tends to the exact charge density as $\mathbf{r}_1 - \mathbf{r}_{1'}$ goes to zero. Thus, the exact KS matrix has exact asymptotics, in contrast with the HF matrix.

The properties discussed above are exemplified in fig. 4, for the H$^-$ ion. Fig. 4 shows contours of the function $8\pi^2 R^2 s^2 Sin(\theta) \Delta\tilde{\rho}(\mathbf{R},\mathbf{s})$, with $\Delta\tilde{\rho} = \tilde{\rho} - \tilde{\rho}_{HF}$ (fig. 4, left-hand part) and $\Delta\tilde{\rho} = \tilde{\rho} - \tilde{\rho}_{KS}$ (fig. 4, right-hand part). Here, the **R/s** representation of the 1RDM is chosen as it has more straightforward connections to **P**-space. In particular, the term $8\pi^2 R^2 s^2 Sin(\theta)$ was picked-up from the six-dimensional term $d\mathbf{R}\,d\mathbf{s}$ appearing in eq. (6); here reduced by symmetry to the three dimensions spanned by $R$, $s$ and $\theta$ (that is, the angle between **R** and **s**, which was set to $\theta = \pi/4$ in fig. 4). This representation makes it possible to appreciate the long-range values of $\Delta\tilde{\rho}$, which substantially impact the computation of the momentum density $\Delta n$.

As expected from the above discussion, the difference $8\pi^2 R^2 s^2 Sin(\theta)(\tilde{\rho} - \tilde{\rho}_{KS})$ is negative everywhere; it has a small spatial extension and a simple nodal structure compared with its HF counterpart. In turn, the difference $B - B_{KS}$ logically has a simpler nodal structure than $B - B_{HF}$ (fig. 5). As a result, the difference $\Delta n = n - n_{HF}$ is more adversely affected than $\Delta n = n - n_{KS}$ at small momenta.

### 3.2 *N*-electron atomic systems

*N*-electron atomic systems are shortly discussed here for the sake of completeness. Some of the properties discussed above can straightforwardly be generalized to a *N*-electron system.

In particular, the exact KS matrix tends advantageously to the exact density as $\mathbf{s} = \mathbf{r}_1 - \mathbf{r}_1'$ goes to zero, so that $\lim_{\mathbf{r}_1' \to \mathbf{r}_1}\{\rho_{KS}(\mathbf{r}_1,\mathbf{r}_1')\} = \rho(\mathbf{r}_1)$. The exact KS matrix should thus be more accurate than the HF matrix at very small **s** values, by construction. However, after projection in the **s** subspace (after integration over **R** = (**r** + **r'**)/2), said advantage is at least partially lost since both $B_{KS}(s)$ and $B_{HF}(s)$ are normalized and have the same limit $B_{KS}(0) = B_{HF}(0) = B(0) = N$.

The extent of this possible improvement can be measured through the kinetic energy $T$. Indeed, since the convexity of the (spherically averaged) auto correlation function $B(s)$ near $s = 0$ is proportional to the average kinetic energy, e.g. $B(s)|_{s \to 0} \approx N - T s^2/3$, the better the latter, the better the auto correlation function at small $s$ (and somehow the momentum density at large momenta, see eq. (6)).

How exactly the charge density affects the momentum density at high momenta can be seen from the formula of Benesch and Smith [32], which have shown that an atomic charge density satisfying Kato's cusp conditions should correspond to a momentum density asymptotically decaying as

$$n(p) \propto \left[Z\rho(0)p^{-4} + O(p^{-6})\right]^2.$$

Accordingly, a momentum density obtained from a determinant giving the exact density (for instance a KS determinant) should be more accurate at large momenta. However, the improvement of KS over HF reflects only the superiority of the exact charge density over the HF density in this case. Accordingly, the key point in this case is therefore the charge density recovered from KS orbitals, as regards coulombic systems.

Next, as regards now large $\mathbf{r}_1$, $\mathbf{r}_{1'}$ values, determinantal 1RDMs are sometimes assumed to tend to (subjected to the limitations outlined in ref. [33]):

$$\lim_{\mathbf{r}_1, \mathbf{r}_1' \to \infty}\{\rho(\mathbf{r}_1,\mathbf{r}_1')\} \approx \rho(\mathbf{r}_1)^{1/2} \rho(\mathbf{r}_1')^{1/2} . \tag{22}$$

Hence, provided that eq. (22) applies, one has the limits:

$$\lim_{\mathbf{r}_1, \mathbf{r}_1' \to \infty} \{\rho_{KS}(\mathbf{r}_1, \mathbf{r}_1')\} \approx \rho(\mathbf{r}_1)^{1/2} \rho(\mathbf{r}_1')^{1/2} \quad \text{and}$$

$$\lim_{\mathbf{r}_1, \mathbf{r}_1' \to \infty} \{\rho_{HF}(\mathbf{r}_1, \mathbf{r}_1')\} \approx \rho_{HF}(\mathbf{r}_1)^{1/2} \rho_{HF}(\mathbf{r}_1')^{1/2} \quad ,$$

which again favors the exact KS matrix.

Should however the KS matrix have an exact asymptotic at large $\mathbf{r}_1$, $\mathbf{r}_1'$, it is not certain that the exact KS momentum density be more accurate than HF at small momenta, as seen for He, in table 1 (see the values obtained from ref. [15], considering that exact KS orbitals are obtained in ref. [15]).

To summarize, the structure of exact KS and exact HF density matrices should look much the same as they derive each from a single determinant resulting in similar charge densities. Thus, whether direct KS momentum densities of coulombic systems improve over HF depends on the quality of the **R**-space density recovered from KS orbitals, as reflects in the results listed in table 1. Another factor will be discussed in the concluding remarks.

### 3.2 Simple corrections to *N*-electron atomic systems

Finally, as outlined above, a major inconsistency of direct KS momentum densities arises due to the fact that said densities do not reflect the true DFT kinetic energy, as part of it resides outside the KS determinant. In this respect, let us remark that $n_{KS}(p)$ may be corrected in a very simple manner so as to be consistent with the exact DFT kinetic energies, e.g. under scaling

$$n_{KS}(p) \to \kappa^3 n_{KS}(\kappa p),$$

leading to scaled momentum moments

$$\langle p^m \rangle_\kappa = \kappa^{-m} \langle p^m \rangle,$$

where $\kappa$ is chosen so as to render the exact *or any guess to the exact* DFT kinetic energy. Notice that amongst the cases studied above, the true $n(p)$ might be obtained by scaling $n_{KS}(p)$ in the case of the Moshinsky atom only.

Though no scheme exists in practice to extract the correlation part of the DFT kinetic energy, the latter can in many cases be inferred. A brute force method is to assume the DFT energy obtained for some system at equilibrium to be the exact one, whereby the kinetic energy follows directly from the virial theorem. Another way is to compute it from suitable functionals, see for example refs. [34,35] and more generally ref. [5].

Then, once the exact or any guess to the exact DFT kinetic energy is known, the above scale factors can be computed so as to recover said kinetic energy from $n_{KS}(p)$. For example, considering the momentum moments obtained from ref. [15] and taking $\kappa = 0.9937$, $0.9975$ and $0.9987$ respectively for He, Be and Ne allows the exact kinetic energy to be recovered. Applying the above scaling further reduces the mean absolute relative error on the reported momentum moments by a factor two on average, that is, reduces it from 0.8% to 0.4%. In comparison, the HF theory leads to an average error of 1.1% for the same systems.

## 4. Conclusion

The question of how DFT compares with HF for the computation of momentum-space properties was addressed in relation to systems for which (near) exact, exact KS and HF one-electron matrices are known. In particular, we considered the Moshinsky atom, the Hooke's atom and light two-electron ions, for which expressions for correlated density-matrices or momentum densities have been derived in closed-form. It appears that using exact or near exact KS orbitals slightly improves over HF when particles (electrons) interact via the Coulomb potential (Hooke's atom, light two-electron ions). This is not true in the case of the Moshinsky atom, where particles repel each other via the Hooke's law.

Such improvements, if any, reflect the quality of the one-electron matrix itself and merely result from the fact the exact KS orbitals render an exact **R**-space density. Though said improvements are somehow artificial, they show that direct KS momentum densities may be more accurate than HF's, even without corrections applied.

As regards now "approximate" DFT and reverting merely to previous studies: whether direct KS momentum densities outperform HF strongly depends on the choice of functionals. However, for the purpose of computing momentum densities or related distributions in a DFT context, said functionals should be carefully chosen, as previously suggested [13, 15]. In particular, it seems necessary to rethink how to implement the specific treatment of correlation (possibly better to skip it, in view of table 1), since the structure of the KS-1RDM does not reflect it.

This is illustrated in the following example. If correlation is specifically incorporated via some functionnals, care should be taken to suitably correct the KS-1RDM, as illustrated in ref. [35]. Indeed, consider for example a Colle-Salvetti's like embodiment of correlation [36], i.e. starting from the following trial correlated wave function

$$\psi = \psi^{(0)}(\mathbf{x}_1, \mathbf{x}_2, ..., \mathbf{x}_N) \prod_{i<j} (1 - \varphi(\mathbf{r}_i, \mathbf{r}_j)),$$

where $\varphi(\mathbf{r}_i, \mathbf{r}_j)$ (assumed real) correlates the electron pair $(i, j)$ and $\psi^{(0)}$ denotes a determinantal approximation to the exact ground-state wavefunction. The 2RDM must develop as

$$\rho_2(\mathbf{r}_1, \mathbf{r}_2; \mathbf{r}_1', \mathbf{r}_2')$$
$$= \rho_2^{(0)}(\mathbf{r}_1, \mathbf{r}_2; \mathbf{r}_1', \mathbf{r}_2')(1 - \varphi(\mathbf{r}_1, \mathbf{r}_2) - \varphi(\mathbf{r}_1', \mathbf{r}_2') + \varphi(\mathbf{r}_1, \mathbf{r}_2)\varphi(\mathbf{r}_1', \mathbf{r}_2')) + R(\mathbf{r}_1, \mathbf{r}_2; \mathbf{r}_1', \mathbf{r}_2'),$$

where $\rho_2^{(0)}$ is the determinantal 2RDM and $R$ includes all the terms that can not be factorized as $\rho_2^{(0)}$ times a correlation factor, i.e. integrals involving $\rho_2^{(0)}$ or higher-order RDMs times correlation factors.

A direct calculation of the 1RDM (applying eq. (2 - 4)) from the above wavefunction shows that it develops as

$$\rho_1(\mathbf{r}_1;\mathbf{r}_1') = \rho_1^{(0)}(\mathbf{r}_1;\mathbf{r}_1')$$
$$+ 2\int \left(-\varphi(\mathbf{r}_1,\mathbf{r}_2) - \varphi(\mathbf{r}_1',\mathbf{r}_2) + \varphi(\mathbf{r}_1,\mathbf{r}_2)\varphi(\mathbf{r}_1',\mathbf{r}_2)\right)\rho_2^{(0)}(\mathbf{r}_1,\mathbf{r}_2;\mathbf{r}_1',\mathbf{r}_2)d\mathbf{r}_2 + ...$$

Thus, the 1RDM must therefore be somehow affected by correlation, contrasting in spirit with [36], where it is assumed that $\rho_1(\mathbf{r}_1;\mathbf{r}_1') \approx \rho_1^{(0)}(\mathbf{r}_1;\mathbf{r}_1')$.

Thus, including correlation within a SCF calculation based on the 2RDM only while ignoring effects on the 1RDM, amounts to create some discrepancies between the fictious system represented by the KS-1RDM and the true 1RDM. Therefore, one understands that inclusion of correlation within a KS-DFT may adversely affect the KS-1RDM, as regards its direct conversion to a momentum density, as outlined in ref. [13].

## Acknowledgments

I am pleased to express my deep gratitude to Andreas Savin, who actually inspired this work, by drawing to my attention critical differences between true and approximate DFT. I further thank him for his interest in this work, his suggestions and interesting discussions on the solutions developed in DFT for momentum-space studies.

TABLE CAPTIONS

Table 1: Comparison of DFT-, HF- and correlated wavefunction-based momentum moments obtained in literature for He, Be and Ne. The percent errors are relative to the correlated wavefunction results.

Table 2: Coefficients in the development, $N = \langle p^0 \rangle$ and $T = \frac{1}{2}\langle p^2 \rangle$ momentum moments of the momentum density, eq. (13), of the Hooke's atom for k = ¼.

Table 3: Comparison between quasi-exact and HF momentum moments of H-, He and the Hooke's atom.

TABLES

Table 1, S. Ragot, Journal of Chemical Physics

| | *He* | | | | | | | | | |
|---|---|---|---|---|---|---|---|---|---|---|
| | *BLYP* [13] | | *PW91* [13] | | *SIC-LDA* [13] | | *Constrained KS orbitals* [15] | | *HF* [18] | | *Correlated WF* [19] |
| $\langle p^{-2}\rangle$ | 4.417 | 7.8 | 4.404 | 7.4 | 3.916 | -4.5 | 4.132 | 0.8 | 4.092 | -0.2 | 4.099 |
| $\langle p^{-1}\rangle$ | 2.208 | 3.3 | 2.206 | 3.2 | 2.104 | -1.6 | 2.149 | 0.5 | 2.141 | 0.1 | 2.139 |
| $\langle p^{1}\rangle$ | 2.775 | -1.4 | 2.773 | -1.5 | 2.834 | 0.7 | 2.797 | -0.6 | 2.799 | -0.6 | 2.815 |
| $\langle p^{2}\rangle$ | 5.748 | -1.0 | 5.726 | -1.4 | 5.836 | 0.5 | 5.734 | -1.3 | 5.723 | -1.4 | 5.807 |
| $\langle p^{3}\rangle$ | 18.60 | 1.1 | 18.417 | 0.1 | 18.393 | -0.1 | 18.110 | -1.6 | 17.990 | -2.3 | 18.406 |
| | *Be* | | | | | | | | | |
| | | | | | | | *Constrained KS orbitals* [15] | | *HF* [18] | | *Correlated WF* [20] |
| $\langle p^{-2}\rangle$ | | | | | | | 23.449 | -6.9 | 25.294 | 15.3 | 21.939 |
| $\langle p^{-1}\rangle$ | | | | | | | 6.122 | -3.6 | 6.3185 | 6.9 | 5.909 |
| $\langle p^{1}\rangle$ | | | | | | | 7.468 | 0.9 | 7.4342 | -1.3 | 7.533 |
| $\langle p^{2}\rangle$ | | | | | | | 29.183 | 0.5 | 29.146 | -0.6 | 29.333 |
| $\langle p^{3}\rangle$ | | | | | | | 185.55 | 0.4 | 185.55 | -0.4 | 186.35 |
| | *Ne* | | | | | | | | | |
| | *BLYP* [13] | | *PW91* [13] | | *SIC-LDA* [13] | | *Constrained KS orbitals* [15] | | *HF* [18] | | *Correlated WF* [21] |
| $\langle p^{-2}\rangle$ | 5.7747 | 4.0 | 5.7437 | 3.4 | 5.5369 | 0.3 | 5.583 | 0.5 | 5.4694 | -1.5 | 5.5527 |
| $\langle p^{-1}\rangle$ | 5.5704 | 1.7 | 5.5583 | 1.5 | 5.3589 | 2.2 | 5.497 | 0.3 | 5.4537 | -0.4 | 5.4782 |
| $\langle p^{1}\rangle$ | 35.104 | -0.4 | 35.11 | -0.4 | 35.352 | -0.3 | 35.156 | -0.2 | 35.197 | -0.1 | 35.2412 |
| $\langle p^{2}\rangle$ | 257.38 | -0.1 | 257.27 | -0.2 | 258.32 | -0.2 | 257.183 | -0.2 | 257.09 | -0.3 | 257.751 |
| $\langle p^{3}\rangle$ | 3598.4 | 0.2 | 3595.8 | 0.1 | 3588.5 | 0.1 | 3583.33 | -0.2 | 3584.2 | -0.2 | 3591.5 |

Table 2, S. Ragot, Journal of Chemical Physics

| $m_{max}$ | $Cm_{max}$ | $N = \langle p^0 \rangle$ | $T = \frac{1}{2}\langle p^2 \rangle$ |
|---|---|---|---|
| 0 | 1. | 2.833029 | 1.06239 |
| 1 | $-6.666667 \cdot 10^{-1}$ | 2.073745 | 0.696685 |
| 2 | $-1.333333 \cdot 10^{-1}$ | 2.00832 | 0.661771 |
| 3 | $-3.809524 \cdot 10^{-2}$ | 1.997905 | 0.656692 |
| 4 | $-1.058201 \cdot 10^{-2}$ | 1.996832 | 0.657177 |
| 8 | $-2.489885 \cdot 10^{-5}$ | 1.999073 | 0.661649 |
| 16 | $-3.061856 \cdot 10^{-12}$ | 1.999873 | 0.663651 |
| 32 | $-3.985972 \cdot 10^{-30}$ | 1.999981 | 0.664179 |
| 64 | $-8.873795 \cdot 10^{-75}$ | 1.999997 | 0.664339 |
| 128 | $-1.346499 \cdot 10^{-182}$ | 1.999999 | 0.6643909 |
| 256 | $-5.149269 \cdot 10^{-436}$ | 2. | 0.6644084 |
| : | | : | : |
| $\infty$ | | 2 | 0.664417 |

Table 3, S. Ragot, Journal of Chemical Physics

|  | H⁻ [19,27] | | | He [19,18] | | | Hooke's atom (this work) | | |
| --- | --- | --- | --- | --- | --- | --- | --- | --- | --- |
|  | "Exact" | HF | | "Exact" | HF | | "Exact" | HF | |
| $\langle p^{-2} \rangle$ | 42,900 | 34,571 | 19,4 % | 4,099 | 4,092 | 0,2 % | 9,04965 | 9,29640 | -2,7 % |
| $\langle p^{-1} \rangle$ | 6,446 | 5,999 | 6,9 % | 2,139 | 2,141 | -0,1 % | 3,39636 | 3,44843 | -1,5 % |
| $\langle p^{1} \rangle$ | 1,115 | 1,098 | 1,5 % | 2,815 | 2,799 | 0,6 % | 1,49986 | 1,46947 | 2,0 % |
| $\langle p^{2} \rangle$ | 1,055 | 0,976 | 7,5 % | 5,807 | 5,723 | 1,4 % | 1,3288 | 1,26507 | 4,8 % |
| $\langle p^{3} \rangle$ | 1,658 | 1,458 | 12,1 % | 18,406 | 17,990 | 2,3 % | 1,343 | 1,22546 | 8,7 % |

FIGURE CAPTIONS

Fig. 1: radial momentum densities of the Moshinsky atom ($k = 1/4$, $l = 1/12$): exact (full line), HF (large dashed) and KS (small dashed). Curves integrating to 0: exact - HF (large dashed) and exact – KS (small dashed) radial differences.

Fig. 2: radial momentum densities for the Hooke's atom ($k = 1/4$). Upper fig.: exact (full line, from eq. (13), stopping the expansion at $m = 13$), HF (large dashed) and KS (small dashed). Lower fig.: radial momentum density differences. Expanded exact - HF (large dashed) and expanded exact - KS (small dashed).

Fig. 3: radial momentum densities of H$^-$: near exact (full line, see text), HF (large dashed) and KS (small dashed).

Fig. 4: "radial" density matrix differences for H$^-$: section in a plane ($R$, $s$) with 10 contours ranging from –3 to 3, the grey line representing a zero contour. Left: correlated – HF, right: correlated – KS.

Fig. 5: auto-correlation function differences of H$^-$. near exact - HF (large dashed) and near exact - KS (small dashed).

FIGURES

Figs 1 - 5 sent separately by FTP